\documentclass{PoS}
\usepackage{cite}
\usepackage[T1]{fontenc} 
\usepackage{amsmath,amssymb,graphicx,bbm,mathrsfs,nicefrac}
\usepackage{braket} 
\usepackage{wasysym} 
\usepackage{slashed} 

\def\mcnlo{\textsc{Mc@Nlo}}
\def\powheg{\textsc{Powheg}}

\def\helacnlo{\textsc{Helac-Nlo}}

\def\mcnlo{\textsc{Mc@Nlo}}
\def\powheg{\textsc{Powheg}}
\def\deductor{\textsc{Deductor}}

\title{NLO calculations  matched  with the Nagy-Soper parton shower}

\ShortTitle{NLO calculations  matched  with the Nagy-Soper parton shower}

\author{Michal Czakon, Heribertus B.  Hartanto, Manfred  Kraus, \speaker{Malgorzata Worek}%
  ~\thanks{Preprint number: TTK-16-04}\\
        Institute for Theoretical Particle Physics and Cosmology\\
        RWTH Aachen University\\
        D-52056 Aachen, Germany\\
        E-mails: \\ \email{mczakon@physik.rwth-aachen.de \\
       hartanto@physik.rwth-aachen.de \\ kraus@physik.rwth-aachen.de \\
       worek@physik.rwth-aachen.de}}

\abstract{ 
An MC$@$NLO-like matching of NLO QCD calculations with the Nagy-Soper
parton shower is briefly summarised. Uncertainties and ambiguities
of the matching scheme are shortly discussed.  A few results for the
$pp \to t\bar{t} j +X$ production process at the LHC with $\sqrt{s}=8$
TeV are also shown. All results have been obtained using the Nagy-Soper
parton shower implementation in the \textsc{Deductor} program together with the
\textsc{Helac-NLO} framework. }

\FullConference{12th International Symposium on Radiative Corrections
  (Radcor 2015) and LoopFest XIV (Radiative Corrections for the LHC
  and Future Colliders), \\15-19 June  2015, \\UCLA Department of Physics
  $\&$ Astronomy Los Angeles, CA, USA}

\begin{document}

%
\section{Introduction}
%

After an almost two year shutdown, the Large Hadron Collider (LHC)
started delivering physics data in 2015. This marked the begin of Run
2 at the LHC and opened the path to an even deeper understanding of
the Standard Model (SM) physics and hopefully to new
discoveries. Protons are now collided at the LHC at the record
breaking energy of $13$ TeV, almost double the collision energy of Run
1 ($\sqrt{s}=7,8$ TeV). On the $15^{\rm th}$ of December 2015, the ATLAS
and CMS experiments both reported a number of results using
$\sqrt{s}=13$ TeV proton collision data. Even though the amount of
data on which analyses are based is still limited, experimentalists
have succeeded in producing numerous results for the SM physics,
including new results for the Higgs boson and the top
quark. Additionally, many searches for physics beyond the SM, yielding
many improved limits beyond Run 1, have been presented. In 2016 we
shall wait for more (and updated) results. To describe experimental
data various theoretical tools are needed, among others, Monte Carlo
(MC) generators that are widely used in all experimental
analyses. Such programs need to be improved as data become more
precise. There are various ways to reach better accuracy of
theoretical predictions. In particular, one such possibility is an
increase of the order in the fixed order perturbation expansion.
Next-to-leading order (NLO) calculations can now be performed in a
fully automatic manner. We mention here only the \textsc{Helac-NLO}
framework \cite{Bevilacqua:2011xh}, that will be used in our
studies. The software has recently been used to calculate NLO QCD
corrections to a process with five objects in the final state, i.e.
$W^+W^-b\bar{b}j$\cite{Bevilacqua:2015qha}.  Adding one more order
complicates the picture tremendously.  For next-to-next-to-leading order
(NNLO) calculations, such automatic frameworks are not yet publicly
available. In addition, a list of processes that have been completed
at that order comprises processes with two objects in the final state
only.  The most prominent example being the NNLO calculation for the
$t\bar{t}$ production process at hadron colliders, where two coloured
and massive final state fermions are present
\cite{Baernreuther:2012ws,Czakon:2013goa}. Finally, an important
milestone has been recently reached when a first calculation at the
next-to-next-to-next-to-leading order (${\rm N^3LO}$) in perturbative
QCD for the Higgs production in gluon fusion in the large top-mass
limit has been completed \cite{Anastasiou:2015ema}.  Since for each
order one needs to include more Feynman diagrams both with one
particle more in the final state and with one loop more in the
intermediate state, a formal order-by-order perturbative calculation
is limited to a few orders only. Alternatively, to improve the
theoretical accuracy one can use parton shower (PS) programs.  Parton
shower algorithms approximate higher-order corrections by including
the leading soft and collinear contributions to all orders. These
programs are not able to correctly estimate the radiation of hard
jets. They can, however, generate arbitrarily many jets in the final
states. In order to improve the simulation of hard jets production in
the parton shower, approaches were developed to match parton showers
with fixed order NLO calculations \cite{Frixione:2002ik,Nason:2004rx}
and to merge matched calculations for different jet multiplicities
both at LO and NLO \cite{Mangano:2006rw,Alwall:2007fs,
Lonnblad:2012ix,Hamilton:2012np,Frederix:2012ps,Hoeche:2012yf,
Gehrmann:2012yg,Hoeche:2014qda}. Finally, those
methods have already started to be applied to the NNLO calculations.
Formally, however, in parton showers that are currently being use by
experimental collaborations the resummation of soft and collinear
contributions is only certified to Leading Logarithmic (LL)
accuracy. In practice, many of the NLL contributions are already
included, either by the angular ordering of the parton shower or by an
optimal scale choice for $\alpha_s$. Thus, yet another progress on the
front of theoretical predictions for the LHC can be accomplished by
improving the parton shower itself. This can be done through an
inclusion of color and spin correlations in subsequent parton shower
emissions. An impact of the colour suppressed terms in parton shower
simulations, for the final state parton shower, have already been
studied in \cite{Platzer:2012np}. First results for the $e^+e^-\to 2j$
production at LEP energy have shown that standard LEP-observables,
e.g. event shapes and jet rates, are only slightly affected.  However,
for tailored observables, i.e. observables that are sensitive to soft
(wide angle) splitting, deviations as large as $20\%$ have been
observed. The first step in the direction of the complete algorithm,
which consistently takes into account both spin and color correlations
in the parton emissions for the collision of hadrons, has been done by
Nagy and Soper \cite{Nagy:2007ty,Nagy:2012bt}.  A large part of their
idea has been implemented in the \textsc{Deductor} program
\cite{Nagy:2014mqa}.  Even though the exact colour and spin
correlations are not yet included in the software, it can already
provide results with an extension of the leading colour approximation,
which has been implemented instead.

In this proceedings we shall briefly summarise the concept behind the
Nagy-Soper parton shower. In the next step we shall outline a matching
procedure between the Nagy-Soper parton shower and a fixed order
calculation at NLO \cite{Czakon:2015cla}. To this end, we use the
\textsc{Mc@Nlo} approach \cite{Frixione:2002ik}. When matching NLO
calculations with parton shower programs, one needs to avoid the
double counting of emissions, which on the one hand can be generated
with a jet from the matrix element and on the other hand can appear as
a jet from the parton shower.  The \textsc{Mc@Nlo} method removes
double counting contributions by expanding the parton shower to first
order in $\alpha_s$ and compensating for the terms, which are already
present at fixed order. Finally, a few results for the $pp \to
t\bar{t}j$ production process, which involves non-trivial colour
exchange, massive partons and requires cuts already at the Born level,
will be shown.

%
\section{Nagy-Soper parton shower}
%

In case of a generic process $ a+b \to m$ the expectation value for a
completely inclusive observable $F$ in the Nagy-Soper formalism can be
written as follows
\begin{equation}
  \sigma[F] = \sum_m \frac{1}{m!}\int [d\{p,f\}_m] \bra{\mathcal{M}(
  \{p,f\}_m)}F(\{p,f\}_m)\ket{\mathcal{M}(\{p,f\}_m)} \frac{f_a(\eta_a,
  \mu_F^2)f_b(\eta_b,\mu_F^2) }{4n_c(a) n_c(b) \times \text{flux}}  \,,
  \label{rho_intro}
\end{equation}
where the sum runs over all final state multiplicities. Here,
$[d\{p,f\}_m]$ is the sum of all $m$-particle phase space measures for
different flavour sequences $\{f\}_m$. The factor $1/m!$ is necessary
to account for identical contributions. The parton density functions
evaluated at the momentum fraction $\eta$ and factorization scale
$\mu_F^2$ are denoted by $f_{a/b}(\eta,\mu_F^2)$.  The factor $4$ in
the denominator comes from averaging over initial state spins and
$n_c(i)$  is the colour factor. The quantity  
\begin{equation}
\rho(\{p,f\}_m) \sim \ket{\mathcal{M}(\{p,f\}_m)}\bra{\mathcal{M}
  (\{p,f\}_m)} \,,
\end{equation}
is the quantum density $\rho$ and a matrix element $\mathcal{M}$ can
be viewed as a vector $\ket{\mathcal{M}(\{p,f\}_m)}$ in colour
$\otimes$ spin space and can be resolved into basis vectors
$\ket{\{s\}_m}$ and $\ket{\{c\}_m}$ with complex expansion
coefficients $\mathcal{M}(\{p,f,s,c\}_m)$
\begin{equation}
      \ket{\mathcal{M}(\{p,f\}_m)} = \sum_{\{s\}_m} \sum_{\{c\}_m} 
     \mathcal{M}(\{p,f,s,c\}_m) \ket{\{s\}_m} \otimes \ket{\{c\}_m} 
	= \sum_{\{s,c\}_m}  \mathcal{M}(\{p,f,s,c\}_m) \ket{\{s,c\}_m}\;.
   \label{expandbasis}
\end{equation}
The propagation of the quantum density matrix $\rho$ from some initial
shower time, $t_0$, to some final shower time, $t_F$, is described by the
evolution equation. The initial shower time corresponds to the hard
interaction, whereas the final shower time characterises the physical
scale at which parton emissions cannot be described
perturbatively. The perturbative evolution is described by a unitary
operator $U(t_F,t_0)$ that obeys
\begin{equation}
  \frac{dU(t,t_0)}{dt} =[\mathcal{H}_I(t) - \mathcal{V}(t)]U(t,t_0)\;.
  \label{diffeq}
\end{equation}
Here, $\mathcal{H}_I(t)$ describes resolved emission and
$\mathcal{V}(t)$ the unresolved/virtual one. The latter can be further
decomposed into a color diagonal, $\mathcal{V}_E(t)$, and a color
off-diagonal part, $\mathcal{V}_S(t)$, as follows
\begin{equation}
\mathcal{V}(t) = \mathcal{V}_E(t) + \mathcal{V}_S(t)\;.
\end{equation}
Eq.~\eqref{diffeq} can be solved as
\begin{equation}
U(t,t_0) = N(t,t_0) + \int_{t_0}^t d\tau~ U(t,\tau)\left[\mathcal{H}_I
  (\tau)-\mathcal{V}_S(\tau)\right]N(\tau,t_0) \;, 
\end{equation}
where $N(t,t_0)$ is the Sudakov form factor (a number) defined as
\begin{equation}
  N(t,t_0) = \exp\left(-\int_{t_0}^t d\tau~\mathcal{V}_E(\tau)
  \right)\;.
\end{equation}
In case of a non-trivial colour evolution, the exponentiation of a
non-diagonal matrix is very difficult. Thus, only the colour diagonal
part, $\mathcal{V}_E(t)$, is exponentiated, whereas the off-diagonal
part, $\mathcal{V}_S(t)$, is treated perturbatively in the same way as
$\mathcal{H}_I(t)$. The expectation value of the observable $F$,
including shower effects, is provided via
\begin{equation} \sigma[F] = (F|\rho(t_F)) =
(F|U(t_F,t_0)|\rho(t_0))\;.
\end{equation}
The evolution in the shower time is always ordered in some chosen
kinematic variable to correctly resum leading logarithms of infrared
sensitive quantities. In the Nagy-Soper parton shower the following
variable has been proposed
\begin{equation}
  \Lambda^2_l = \frac{|(\hat{p}_l \pm \hat{p}_{m+1})^2-m^2_l|}
  {2p_l\cdot Q_0} Q_0^2\;, ~~~~~~~~~~~~~~~~~~~~~~
e^{\,-t}=\frac{\Lambda^2_l}{Q_0^2}\,,
  \label{lambdadef}
\end{equation}
where $\hat{p}_l$ is the emitter momentum after emission,
$\hat{p}_{m+1}$ the emitted parton momentum, $p_l$ the emitter
momentum before emission ($p_l^2=m^2_l$) and $Q_0$ is the total final
state momentum. The minus/plus sign between $\hat{p}_l$ and
$\hat{p}_{m+1}$ in Eq.~\eqref{lambdadef} applies to an initial/final
state splitting.

%
\section{Matching Inclusive Processes}
%

There exist several schemes for matching NLO calculations
with parton showers, the most popular being \powheg{} and \mcnlo{}. In
order to benefit from the recently implemented subtraction scheme
based on the Nagy-Soper parton shower splitting kernels
\cite{Bevilacqua:2013iha}, which is implemented in the
\textsc{Helac-Dipoles} framework \cite{Czakon:2009ss}, we chose the
\textsc{Mc@Nlo} formalism. For a generic $a + b \to m$ process at NLO,
one can write the quantum density matrix in a perturbative expansion
in $\alpha_s$, according to
\begin{equation}
  |\rho) = \underbrace{|\rho_m^{(0)})}_{\text{Born} , \; \mathcal{O}(1)} + 
  \underbrace{|\rho_m^{(1)})}_{\text{Virtual}, \; \mathcal{O}(\alpha_s)} + 
  \underbrace{|\rho_{m+1}^{(0)})}_{\text{Real}, \; \mathcal{O}(\alpha_s)} +
  \mathcal{O}(\alpha_s^2)\;,
  \label{rho_defNLO}
\end{equation}
where the leading order contribution is counted as order $1$ in the
strong coupling $\alpha_s$. In addition, $|\rho_m^{(0)})$ and
$|\rho_{m+1}^{(0)})$ are tree level matrix elements, whereas
$|\rho_m^{(1)})$ is the one-loop amplitude.  Based on this quantum
density matrix, the expectation value of the observable $F$ including
shower effects $\sigma [F]^{PS} = (F|U(t_F,t_0)|\rho) $ suffers from
double counting. This can be seen when the iterative solution to the
evolution equation is expanded to $\mathcal{O}(\alpha_s)$. Indeed, we
obtain
\begin{equation}
  |\rho(t_F))= U(t_F,t_0)|\rho) \approx |\rho) + \int_{t_0}^{t_F} d\tau 
  \left[ \mathcal{H}_I(\tau) - \mathcal{V}(\tau)\right]|\rho_m^{(0)})+
  \mathcal{O}(\alpha_s^2)\;,
  \label{PSexpand}
\end{equation}
and $|\rho(t_F))$ contains the first emission contributions twice, once from
the real emission quantum density $|\rho^{(0)}_{m+1})$, and once from
the parton shower approximation $\mathcal{H}_I(\tau)|\rho^{(0)}_m)$.
The problem of double counting can be overcome if the density matrix
is modified  as follows
\begin{equation}
  |\bar{\rho}) \equiv |\rho) - \int_{t_0}^{t_F} d\tau \left[ \mathcal{H}_I
  (\tau) - \mathcal{V}(\tau)\right]|\rho_m^{(0)})+ \mathcal{O}(\alpha_s^2)
  \;.
  \label{RhoBar}
\end{equation}
Considering $U(t_F,t_0)|\bar{\rho})$ and expanding the evolution
equation up to $\mathcal{O}(\alpha_s)$ shows that the undesired parton
shower contributions are cancelled.  Thus, for an infrared safe observable
$F$, we can write 
\begin{equation}
\begin{split}
  \bar{\sigma} [F] &=\frac{1}{m!}\int  [d\Phi_m] (F|U(t_F,t_0) |\Phi_m) 
  \left[  (\Phi_m|\rho_m^{(0)})  + (\Phi_m|\rho_m^{(1)}) + \int_{t_0}^{t_F} 
  d\tau (\Phi_m|\mathcal{V}(\tau)|\rho_m^{(0)})	 \right]  \\
  &+\frac{1}{(m+1)!} \int [d\Phi_{m+1}] (F|U(t_F,t_0) |\Phi_{m+1}) \left[
   (\Phi_{m+1}|\rho_{m+1}^{(0)}) - \int_{t_0}^{t_F} d\tau 
   (\Phi_{m+1}|\mathcal{H}_I(\tau)|\rho_m^{(0)}) \right]\;,
\end{split}
  \label{XsecMatched}
\end{equation}
where $\Phi_\lambda = \{p,f,s^\prime, c^\prime, s,c\}_\lambda$ and
$\lambda$ can be either $m$ or $m+1$. The parton shower splitting
kernels are used to provide subtraction terms for the infrared
singularities
\begin{equation}
  \int_{t_0}^\infty d\tau~\mathcal{H}_I(\tau) = \sum_l \mathbf{S}_l
  \int_0^\infty d\tau~\delta(\tau-t_l)\Theta(\tau - t_0) 
  = \sum_l \mathbf{S}_l \Theta(t_l - t_0)\;,
  \label{H_sum}
\end{equation}
\begin{equation}
  \int_{t_0}^\infty d\tau~ \mathcal{V}(\tau) = \sum_l \int d\Gamma_l\; 
  \mathbf{S}_l \Theta(t_l-t_0) \equiv \mathbf{I}(t_0) + \mathbf{K}(t_0)\;,
  \label{Vtheta}
\end{equation}
where the sum runs over all external legs and $\mathbf{S}_l$ is the
total splitting kernel for a given external leg $l$. The parameter
$t_l$ is the shower time and $\Theta(t_l-t_0)$ represents the ordering
of the emissions. The phase space integration of the additional parton
is denoted by $d\Gamma_l$.  The decomposition of the integrated
$\mathcal{V}(\tau)$ into two operators $\mathbf{I}(t_0)$ and
$\mathbf{K}(t_0)$ is arbitrary. However, we choose $\mathbf{I}(t_0)$
to match the divergencies of the virtual amplitude. In addition, the
limit $t_F\to \infty$ has been taken. This is a source of a mismatch
between the fixed order and the shower calculation, however, it is
numerically small due to the exponential damping by the Sudakov form
factor. The total cross section, which includes parton shower
evolution amounts to
\begin{equation}
  \bar{\sigma} [F]^{PS} =\frac{1}{m!}\int [d\Phi_m]  (F|U(t_F,t_0)|\Phi_m) 
  (\Phi_m|S) 
  + \frac{1}{(m+1)!} \int [d\Phi_{m+1}]  (F|U(t_F,t_0)|\Phi_{m+1}) 
  (\Phi_{m+1}|H)\;,
  \label{Finaleventgeneration}
\end{equation}
where  the following shorthands has been defined 
\begin{align}
	&(\Phi_m|S) \equiv (\Phi_m|\rho_m^{(0)}) + (\Phi_m|\rho_{m}^{(1)})  + 
(\Phi_m|[\mathbf{I}(t_0) + \mathbf{K}(t_0) + \mathbf{P}]|\rho_m^{(0)})\;, 
\label{Sevent0}\\
	&(\Phi_{m+1}|H) \equiv (\Phi_{m+1}|\rho_{m+1}^{(0)}) -  \sum_l 
(\Phi_{m+1}|\mathbf{S}_l|\rho_m^{(0)}) \Theta(t_l-t_0)\;.
\label{Hevent0}
\end{align}
The whole procedure is divided into two steps. First the samples
according to Eq.~\eqref{Sevent0}~and~\eqref{Hevent0} are generated and
afterwards $U(t_F,t_0)$ is applied.

%
\section{Matching Exclusive Processes}
%

For processes with massless partons already at leading order, the
matching prescription as described in the previous section must be slightly
modified by the inclusion of generation cuts. The following replacements
\begin{align}
 \label{Sevent}
 (\Phi_m|S) &\to (\Phi_m|S)F_I(\{\hat{p},\hat{f}\}_m)\;, \\
 \label{Hevent}
 (\Phi_{m+1}|H) &\to (\Phi_{m+1}|H)F_I(\{p,f\}_{m+1})\;,
\end{align}
can be made, where $F_I(\{p,f\}_\lambda)$ is a jet function applied
during the generation of events, on the momenta and flavours of
$\Phi_\lambda$. However, when the parton shower is applied to these
ensembles we can easily see that double counting is still
there. Indeed, after expanding the evolution operator we obtain 
\begin{equation}
\begin{split}
  \bar{\sigma}[F]^{PS} 
  &\approx\frac{1}{m!}\int [d\Phi_m](F|\Phi_m)(\Phi_m|\left[|\rho_m^{(0)})
  +|\rho_m^{(1)}) +\mathbf{P}|\rho^{(0)}_m) \right]
  F_I(\{\hat{p},\hat{f}\}_m) \\
  &+\frac{1}{(m+1)!} \int [d\Phi_{m+1}] (F|\Phi_{m+1})(\Phi_{m+1}|\rho^{(0)}
  _{m+1})F_I(\{p,f\}_{m+1}) \\
  &+\int \frac{[d\Phi_m]}{m!}\frac{[d\Phi_{m+1}]}{(m+1)!}\int_{t_0}^{t_F}
  d\tau~(F|\Phi_{m+1})(\Phi_{m+1}|\mathcal{H}_I(\tau)|\Phi_m)  \\
 &\times(\Phi_m|\rho^{(0)}_m) \Big[F_I(\{\hat{p},\hat{f}\}_m)-
  F_I(\{p,f\}_{m+1})\Big] + \mathcal{O}(\alpha_s^2)
  \;,
  \label{naivejetmatching}
\end{split}
\end{equation}
where the $\mathbf{I}(t_0)+\mathbf{K}(t_0)$ contribution of
$(\Phi_m|S)$ has been cancelled by the linear expansion of the Sudakov
form factor. This mismatch is cured by enforcing the subtraction terms
to fulfill $F_I(\{\hat{p},\hat{f}\}_m)$, i.e.  by modifying the real
subtracted cross section according to
\begin{equation}
  (\Phi_{m+1}|H) \longrightarrow (\Phi_{m+1}|\tilde{H}) 
  \equiv (\Phi_{m+1}|\rho_{m+1}^{(0)}) -  \sum_l 
  (\Phi_{m+1}|\mathbf{S}_l|\rho_m^{(0)}) \Theta(t_l-t_0)
  F_I(Q_l(\{p,f\}_{m+1}))\;,
  \label{modSub}
\end{equation}
where we make use of the inverse momentum mapping $Q_l$ and obtain
\begin{equation}
  F_I(Q_l(\{p,f\}_{m+1})) = F_I(\{\hat{p},\hat{f}\}_m)\;.
\end{equation}
This modification allows us to introduce restrictions on the
functional form of $F_I$.  Expanding the shower
evolution yields
\begin{equation}
\begin{split}
  \bar{\sigma}[F]^{PS} 
  &\approx\frac{1}{m!}\int [d\Phi_m](F|\Phi_m)(\Phi_m|\left[|\rho_m^{(0)})
  +|\rho_m^{(1)}) + \mathbf{P}|\rho^{(0)}_m)\right]
  F_I(\{\hat{p},\hat{f}\}_m) \\
  &+\frac{1}{(m+1)!} \int [d\Phi_{m+1}] (F|\Phi_{m+1})(\Phi_{m+1}|\rho^{(0)}
  _{m+1})F_I(\{p,f\}_{m+1}) \\
  &+\int \frac{[d\Phi_m]}{m!}\frac{[d\Phi_{m+1}]}{(m+1)!}\int_{t_0}^{t_F}
  d\tau~(F|\Phi_{m+1})(\Phi_{m+1}|\mathcal{H}_I(\tau)|\Phi_m)\\
  &\quad\times(\Phi_m|\rho^{(0)}_m) \Big[1-F_I(\{p,f\}_{m+1})\Big]
  F_I(\{\hat{p},\hat{f}\}_m)+\mathcal{O}(\alpha_s^2)
  \;. 
  \label{good_matching}
\end{split}
\end{equation}
The double counting is removed from Eq.~\eqref{good_matching}, 
when the following condition is true 
\begin{equation}
  \Big[1-F_I(\{p,f\}_{m+1})\Big]F(\{p,f\}_{m+1})=0\;.
  \label{mismatch}
\end{equation}
We have used here the fact that $(F|\Phi_{m+1}) \sim
F(\{p,f\}_{m+1})$. This can be achieved when 
\begin{equation}
  F_I(\{p,f\}_{m+1}) = 1\;, \text{ for } F(\{p,f\}_{m+1})\neq 0 \;.
\end{equation}
The interpretation of the last equation is straightforward. Generation
cuts ($F_I(\{p,f\}_{m+1})$) need to be more inclusive than the cuts on
the final observable ($F(\{p,f\}_{m+1})$). Let us note here, that the
\textsc{Mc@Nlo} matching formalism in combination with the Nagy-Soper
parton shower introduces the intrinsic uncertainties. Let us name just
the most important ones here. The Nagy-Soper parton shower treats
bottom and charm quarks as massive. On the other hand in the NLO
calculation we treat them as massless. Thus during the matching
procedure masses for the relevant quarks are introduced by the
on-shell projection. In addition, PDFs are evolved differently in the
NLO calculation and in the shower. In the former case, NLO PDFs are
used, whereas in the latter PDFs are evolved using Nagy-Soper
splitting kernels. The evolution is, however, of higher order and the
NLO accuracy is maintained if the evolutions share a common point
e.g. at the low scale. Finally, the choice of the initial parton
shower time $t_0$ in the parton evolution is arbitrary. We only
require that NLO prediction is recovered for hard emissions.  Thus,
different choices of $t_0$ that fulfil this condition are possible. In
consequence, the $t_0$ quantity should be varied to study the
uncertainty of NLO+PS matching.

%
\section{Implementation}
%

The matching scheme has been implemented within the \helacnlo{}
framework~\cite{Bevilacqua:2011xh}.  We have used it in conjunction
with the Nagy-Soper parton shower present in \deductor{} version
1.0.0~\cite{Nagy:2014mqa}.  Since \deductor{} uses spin averaged
splitting functions, we only provide unpolarized event samples for
showering.  We supply only leading colour events even though
\deductor{} works with the $LC+$ approximation, which comprises a full
color description for collinear and soft-collinear limits and the $LC$
one for pure soft limits. The \textsc{Helac-Nlo} Monte Carlo program
generates an event sample ready for showering with \textsc{Deductor}.
We produce events subprocess by subprocess. First, we use
\textsc{Helac-1Loop} \cite{vanHameren:2009dr} to obtain a set of
unweighted leading order events with the virtual contributions. This
set of weighted events is subsequently reweighted using
\textsc{Helac-Dipoles} in order to include the parton shower virtual
operator, which corresponds to taking into account the integrated
subtraction terms. The real radiation events are generated separately
with \textsc{Helac-Dipoles}. The program has been extended to provide
unweighted events with positive and negative weights, which is
possible, because both the real radiation and the respective
subtraction weights correspond to the same phase space point. For each
accepted event, we choose the most probable diagonal colour flow
configuration. The generated events are stored in a Les Houches file
and transferred to \textsc{Deductor}, which requires an on-shell
projection for charm and bottom quarks and a translation of the color
flow language from the LHE file to internal representation in 
\textsc{Deductor} in terms of color strings.

%
\section{Top quark pair production in association with a jet at
  NLO+PS}
%

We present now the results for the $pp \to t\bar{t} j$ production at
the LHC with $\sqrt{s}= 8$ TeV. We use stable top quarks and put $m_t
= 173.5$ GeV.  The charm and bottom quarks are considered to be
massless at fixed order. Results are obtained using the
\textsc{Mstw2008nlo} PDF set~\cite{Martin:2009iq} with five active
flavours and the corresponding two-loop running of the strong
coupling. The renormalization and factorisation scales are set to
$m_t$, and the starting shower time to
\begin{equation} e^{-t_0} = \min_{i\neq j}\left\{\frac{2 p_i \cdot
p_j} {\mu_T^2 Q_0^2}\right\}\;,
  \label{eq:T0}
\end{equation}
where $p_i$ and $p_j$ are external momenta, $Q_0$ is the total final
state momentum and $\mu_T=1$ for the central prediction.  The
anti-$k_T$ jet algorithm with the separation parameter $R=1$ is used
to cluster partons with pseudorapidity $|\eta| < 5$. The resulting
jets are ordered according to their $p_T$. We also require the tagged
jets to have transverse momentum of $p_T > 50$ GeV and rapidity in the
range of $|y| < 5$. Our analysis is restricted to the perturbative
parton shower evolution. Decays of unstable particles, hadronization
and multiple interactions are not taken into account. The parton
shower treats the charm and bottom as massive particles, thus, 
we use $m_c = 1.4$ GeV and $m_b = 4.75$ GeV. Additionally, the
\textsc{Mstw2008nlo} PDF set at $\mu_F = 1$ GeV is provided as the
starting point for the evolution in \textsc{Deductor}. We also use the
corresponding two-loop running of $\alpha_s$, and restrict the parton
shower to the leading colour approximation.  Results presented in the
following are accurate up to ${\cal O}(1/N_c^2)$.

\begin{figure}[t!]
\begin{center}
  \includegraphics[width=0.49\textwidth]{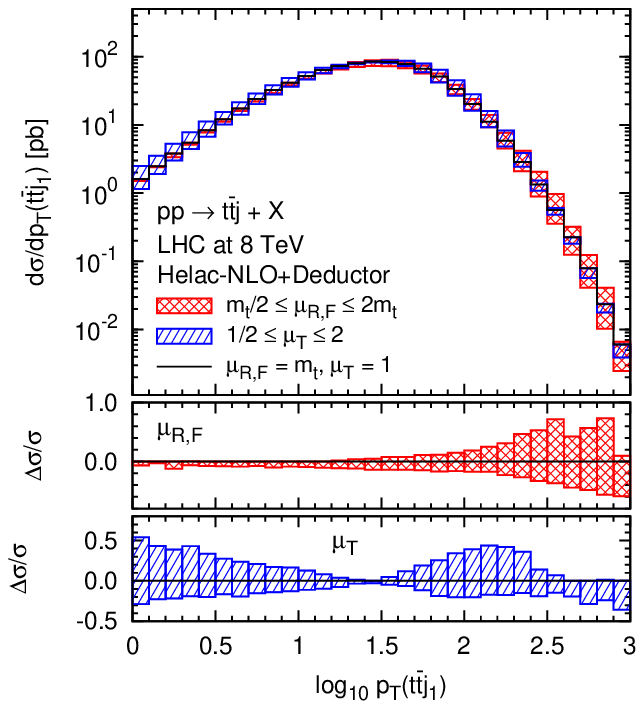}
  \includegraphics[width=0.49\textwidth]{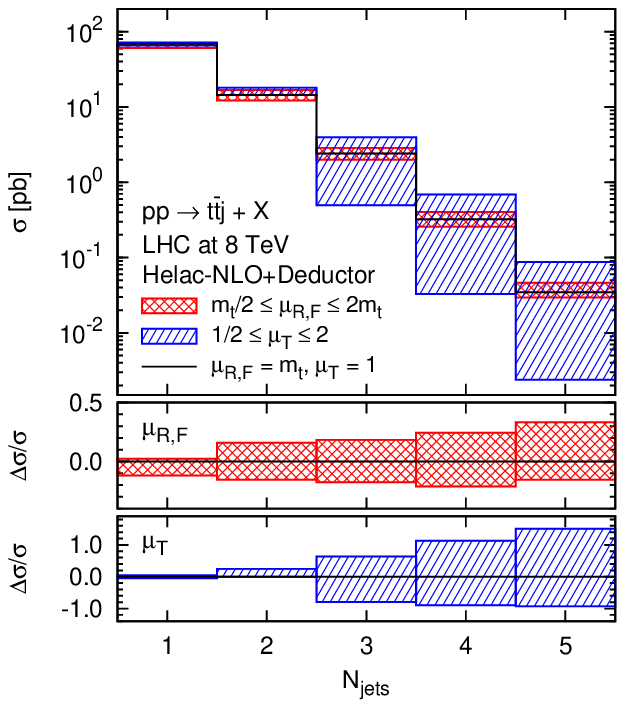}\\
\end{center}
  \caption{\textit{ Differential cross section distributions as a
      function of the transverse momentum of the $t\bar{t}j_1$  system
      (left panel) and inclusive jet cross sections (right panel) 
       for $pp\to t\bar{t}j +X$ at the LHC with $\sqrt{s} = 8$ TeV. Results are
      produced by matching \textsc{Helac-Nlo} predictions to
      \textsc{Deductor}.  The uncertainty bands depict scale and
      initial shower time variation.  The lower panels display the
      corresponding relative deviation from the central value,
      separately for $\mu_{R,F}$ and $\mu_T$.}} 
  \label{fig:pT_scalevar}
\end{figure}

In Fig.~\ref{fig:pT_scalevar} we present the transverse momentum of
the $t\bar{t}j_1$ system together with inclusive jet cross sections.
%
%
The variation bands for $\mu_{R,F}$ and $\mu_T$ have been obtained
using the following sets of three parameter values: $\mu_{R,F}=\{
m_t/2, m_t, 2m_t \}$ and $\mu_T = \{ 1/2,~1,~2 \,\}$,
respectively. The lower panels display corresponding relative
deviations from the central value, separately for $\mu_{R,F}$ and
$\mu_T$. We start the discussion with the transverse momentum of the
$p_T(t\bar{t}j_1)$ system, that is given in the left panel of
Fig.~\ref{fig:pT_scalevar}.  At LO the transverse momentum of this
system is zero due to momentum conservation. When NLO contributions
are included, this observable diverges as the transverse momentum of
the entire system goes to zero. Thus, it can only be reliably
described by the fixed order calculation  in the high $p_T$
region. The situation is changed when the parton shower is
included. In that case, the low $p_T$ behaviour is altered strongly by
the Sudakov form factor as can be seen in Fig.~\ref{fig:pT_scalevar}.
We observe a moderate dependence on $\mu_T$ in the low $p_T$ region,
up to a factor of $1.5$ at the lower end of the spectrum.  This
dependence decreases down to just a few percent around $30$ GeV,
whereas for moderate values of $p_T(t\bar{t}j_1)$ it is at the level
of $20\%-45\%$.  The behaviour  is reversed for the renormalization and
factorisation scale dependence. Visible deviations from the central
value occur once the matrix element dominates and grow substantially
up to almost $80\%$ at the end of the spectrum.

The inclusive jet cross sections are given in the right panel of
Fig.~\ref{fig:pT_scalevar}.  The NLO cross section with exactly one
jet, which is given in the first bin, is rather insensitive to
$\mu_T$, i.e. at the $12\%$ level.  The $\mu_T$ dependence is slightly
larger in the second bin, where the two jet cross section, correct
only at the LO level, is stored.  Starting from the third bin, cross
sections are described via the shower evolution alone, thus, we
observe fairly large variations for both parameters, $\mu_T$ and
$\mu_{R,F}$.  For example, the scale dependence for the cross section
with five jets is found to be around $35\%$.

%
\section{Summary}
%

To summarise, we have outlined a NLO matching scheme for the
Nagy-Soper parton shower. We based our construction on the original
\textsc{Mc@Nlo} approach. We have presented the general formulation
and a few results for top-quark pair production in association with a
jet at the LHC. All results given here have been obtained using an
implementation within the framework of the public codes
\textsc{Helac-Nlo} and \textsc{Deductor}.

%
\section*{Acknowledgement}
%

This research was supported in part by the German Research Foundation (DFG)
under grant WO 1900/1-1 $-$ {\it Signals and Backgrounds Beyond Leading
Order. Phenomenological studies for the LHC}.

\end{document}